\documentclass[twocolumn]{aa}

\usepackage{graphicx}
\usepackage{txfonts}
\usepackage{color}

\begin{document}

\title{On the influence of high energy electron populations on metal abundance
estimates in galaxy groups and clusters.}

\author{D. A. Prokhorov\inst{1,2}}

\offprints{D.A. Prokhorov \email{prokhoro@iap.fr}}

\institute{Institut d'Astrophysique de Paris, CNRS, UMR 7095,
Universit\'{e} Pierre et Marie Curie, 98bis Bd Arago, F-75014
Paris, France
            \and
            Moscow Institute of Physics and Technology,
            Institutskii lane, 141700 Moscow Region, Dolgoprudnii, Russia
            }

\date{Accepted . Received ; Draft printed: \today}

\authorrunning{D.A. Prokhorov}

\titlerunning{Metal abundance estimates}

\abstract
{}
{Spectral line emissivities have usually been calculated for a
Maxwellian electron distribution. But many theoretical works on
galaxy groups and clusters and on the solar corona suggest to
consider modified Maxwellian electron distribution functions to
fit observed X-ray spectra. Here we examine the influence of high
energy electron populations on measurements of metal abundances. }
{A generalized approach which was proposed in the paper by
Prokhorov et al. (2009) is used to calculate the line emissivities
for a modified Maxwellian distribution. We study metal abundances
in galaxy groups and clusters where hard X-ray excess emission was
observed. }
{We found that for modified Maxwellian distributions the argon
abundance decreases for the HCG 62 group, the iron abundance
decreases for the Centaurus cluster and the oxygen abundance
decreases for the solar corona with respect to the case of a
Maxwellian distribution. Therefore, metal abundance measurements
are a promising tool to test the presence of high energy electron
populations.}
{}


\maketitle

\section{Introduction}

Galaxy clusters are large structures in the Universe, with radii
of the order of a megaparsec. Groups of galaxies are the poorest
class of galaxy clusters. The space between galaxies in clusters
is filled with low-density $\sim 10^{-3}$ cm$^{-3}$ high
temperature ($k_{\mathrm{B}} T\sim 1-10$ keV) gas (for a review,
e.g. Sarazin 1986). The temperatures of 1-10 keV are close to the
values of the K-shell ionization potentials ($I_{\mathrm{Z}}=Z^2
Ry$, where Z is the atomic number and Ry is the Rydberg constant)
of heavy elements with atomic numbers in the range of Z=10-26.

Emission lines from heavy elements were detected by X-ray
telescopes from galaxy groups and clusters. The current
instruments (XMM-Newton, Chandra and Suzaku) have provided precise
measurements of the chemical abundances of many elements (O, Ne,
Mg, Si, C, Ar, Ca, Fe and Ni) in groups and clusters. Metal
abundances around 0.5 in Solar units of Anders \& Grevesse (1989)
are derived under the assumptions of collisional equilibrium (for
a review, see Werner et al. 2008).

The ionization rates, recombination rates and emissivity in a
spectral line have usually been calculated for a Maxwellian
electron distribution (e.g. Mewe \& Gronenschild 1981). However,
in many low-density astrophysical plasmas, the electron
distribution may differ from a Maxwellian distribution (e.g.
Porquet et al. 2001).

Hard X-ray tails reported from BeppoSAX observations in the X-ray
spectra of some galaxy clusters (Fusco-Femiano et al. 1999;
Fusco-Femiano et al. 2004; Rossetti \& Molendi 2004 for the Coma
cluster; Kaastra et al. 1999 for the Abell 2199; Molendi et al. 2002
for the Centaurus cluster) were interpreted as bremsstrahlung
emission from non-thermal subrelativistic electrons (see e.g.
Sarazin \& Kempner 2000) or from thermal electrons with a Maxwellian
spectrum distorted by a particle acceleration mechanism (Blasi 2000;
Liang et al. 2002, Dogiel et al. 2007). The bremsstrahlung
interpretation is associated with a huge energy output of emitting
particles and faces energetics problems (e.g. Petrosian 2001).
Evidence for a hard X-ray excess above the thermal emission was also
discovered in galaxy groups with ASCA (Fukazawa et al. 2001,
Nakazawa et al. 2007). The evidence for and the nature of hard X-ray
spectral tails in these galaxy groups and clusters are discussed in
the review by Rephaeli et al. (2008).

The use of non-extensive thermo-statistics (Tsallis 1988; for a
review, see Tsallis 1999), based on the natural generalization of
entropy for systems with long-range interactions, was proposed by
Hansen (2005) to fit the X-ray spectrum observed near NGC 4874
(near the center of the Coma cluster). We consider non-extensive
thermo-statistics as another approach to explain hard X-ray excess
in groups and clusters in the framework of the bremsstrahlung
model.

A more traditional interpretation of hard X-ray tails based on the
inverse Compton scattering (ICS) of relativistic electrons on
relic photons (Sarazin \& Lieu 1998) faces a serious problem. The
combination of hard X-ray and radio observations within the ICS
model implies a magnetic field
%
%
much lower than the values derived from Faraday rotation
measurements (e.g. Clarke et al. 2001). Yet several arguments have
been proposed to alleviate (at least in part) the disagreement (for
a review, see Brunetti 2003; Ferrari et al. 2008; Petrosian et al.
2008).

The presence of high energy subrelativistic electrons (non-thermal
subrelativistic electrons or thermal electrons with a Maxwellian
spectrum distorted by the particle acceleration mechanism) or the
use of non-extensive thermo-statistics must be probed using
various observational methods in order to test interpretations of
X-ray tails from galaxy clusters.

The Sunyaev-Zel'dovich (SZ) effect can be used to constrain the
electron distribution in galaxy clusters. The study of the
influence of high energy subrelativistic electrons on the SZ
effect was done for the Coma and Abell 2199 clusters by Blasi et
al. (2000) and Shimon \& Rephaeli (2002). A method based on the
measurement of the spectral slope around the crossover frequency
of the SZ effect was proposed by Colafrancesco et al. (2009) to
analyse the high energy electron populations in galaxy clusters.

A new probe to study the electron distribution in galaxy clusters,
namely the flux ratio of the emission lines due to FeK$\alpha$
transitions (FeXXV and FeXXVI) was considered by Prokhorov et al.
(2009). This flux ratio is very sensitive to the population of
electrons with energies higher than the ionization potential of a
FeXXV ion (which is $\approx$ 8.8 keV). The influence of the high
energy subrelativistic electron population on the flux ratio is
more prominent in low temperature clusters (as Abell 2199) than in
high temperature clusters (as Coma), because the fraction of
thermal electrons with energies higher than the helium-like iron
ionization potential in low temperature clusters is smaller than
that in high temperature clusters. However, the FeXXVI line is
weak in low temperature clusters and current instruments do not
have sufficient sensitivity to measure the iron line flux ratio.

Kaastra et al. (2009) have shown that relative intensities of the
satellite lines are sensitive to the presence of supra-thermal
electrons in galaxy clusters and that the instruments on future
missions like Astro-H and IXO will be able to demonstrate the
presence or absence of these supra-thermal electrons.

In this paper we study the influence of high energy electron
populations on metal abundance estimates in galaxy groups and
clusters and show that the effect of high energy particles can be
significant. This effect is a promising test to the presence of
high energy subrelativitic electrons in galaxy groups and clusters
because of substantial changes in abundance estimates for modified
Maxwellian distributions. We also consider the effect of high
energy electrons on abundance estimates in the solar corona where
the presence of modified Maxwellian electron distributions has
been proposed.

The paper is organized as follows. In Sect.~2.1 we choose a galaxy
group and a galaxy cluster where high energy subrelativistic
electron populations have been proposed and derive values of the
electron distribution parameters. We calculate the changes in
metal abundances with respect to the values for a Maxwellian
distribution in Sect.~2.2. We discuss the bremsstrahlung model of
hard X-ray emission from galaxy clusters in Sect.~3 and present
our conclusions in Sect.~4. We calculate an oxygen abundance drop
for the solar corona in Appendix A.

\section{Metal abundances in groups and clusters with a high energy electron
population.}

Usually metal abundances are derived under the assumption of a
Maxwellian electron distribution. Let us consider here the
influence of high energy subrelativistic electron populations on
metal abundance determinations.

\subsection{High energy subrelativistic electron populations in galaxy groups and clusters.}

Since we want to analyse the influence of a high energy
subrelativistic electron population on abundance estimates of
chemical elements with atomic numbers Z$\leqslant$26, we must
consider cool clusters where the influence of high energy
subrelativistic electrons on impact excitation and ionization is
more important.

The two objects which will be considered below are the HCG 62 group
and the Centaurus cluster, with respective temperatures of 1 keV and
3.5 keV. These objects are interesting because of hard X-ray excess
detections by Fukazawa et al. (2001) and Molendi et al. (2002),
suggesting a possible high energy subrelativistic electron component
if these hard X-ray excesses are interpreted via bremsstrahlung
emission.

The HCG 62 group is a bright group of galaxies at a redshift
z=0.0146. The best fit temperature is kT=$0.95\pm 0.03$ keV in the
energy band below 2.5 keV (Nakazawa et al. 2007). A hard X-ray
excess from this galaxy group was discovered by Fukazawa et
al. (2001). The highly significant hard X-ray signal in the energy
band 4.0-8.0 keV, of which only $\sim 25\%$ can be accounted for by
thermal IGM (intragalactic medium) emission, was reconfirmed by
Nakazawa et al. (2007). Abundances of Mg, Si, S and Fe were obtained
with Suzaku by Tokoi et al. (2008).

The Centaurus cluster (Abell 3526) is amongst the nearest (z=0.0114)
and brightest clusters in the X-ray sky. Its average gas temperature
is kT=$3.6\pm 0.1$ keV (Molendi et al. 2002). Molendi et al. (2002)
detected a hard X-ray excess at the 3.6$\sigma$ level and concluded
that it is impossible from the Beppo-SAX PDS data alone to establish
the origin of this emission. The abundances of chemical elements in
the Centaurus cluster were studied by Molendi et al. (2002) and
Fabian et al. (2005).

To interpret hard X-ray spectral tails in the framework of the
bremsstrahlung model, different electron distributions were
proposed (e.g. Dogiel 2000; Sarazin \& Kempner 2000; Dogiel et al.
2007). In the paper by Sarazin \& Kempner (2000) it is assumed
that the supra-thermal electron populations start at an electron
kinetic energy 3kT, where T is the temperature of the intracluster
medium (ICM). This electron distribution was considered by Shimon
\& Rephaeli (2002) for an analysis of the influence of
supra-thermal electrons on the SZ effect and by Prokhorov et
al. (2009) for an analysis of electron distributions by means of
the flux ratio of iron lines FeXXV and FeXXVI. In this case the
electron distribution function is given by:
\begin{eqnarray}
&&f_{\mathrm{1}}(x)=f_{\mathrm{M}}(x), \ \ \ \, x<3\\ \nonumber
&&f_{\mathrm{1}}(x)=f_{\mathrm{M}}(x)+\lambda x^{-(\mu+1)/2}, \ \
\ \, x\geq3
\end{eqnarray}
where x=E/kT, $f_{\mathrm{M}}(x)$ is a Maxwellian function,
$\mu=3.33$ is taken from Sarazin \& Kempner (2000) and the
normalization coefficient $\lambda$ is calculated from
observational data. For the calculations of the ionization,
excitation and recombination rates the electrons with very high
energies ($\geq$20 kT) have negligible effect (Porquet et al.
2001), therefore we can place the cut-off at any energy above that
of 20 kT without changing line emissivities.

Another approach to fit the X-ray spectra of galaxy clusters in
the framework of the bremsstrahlung model was proposed by Hansen
(2005). He considered the ICM in thermodynamical equilibrium, but
with an electron distribution function which is defined through
non-extensive thermo-statistics (Tsallis 1988). Reasons for using
Tsallis statistics in galaxy clusters are discussed in Sect.~4 of
Hansen (2005). The equilibrium distribution function in
non-extensive thermo-statistics (e.g. Silva et al. 1998) is:
\begin{equation}
f_{\mathrm{2}}(x)=C \sqrt{x} \left(1-(q-1) x\right)^{1/(q-1)}
\end{equation}
where C is the normalization constant, and q is the parameter
quantifying the degree of non-extensivity.

The electron distribution $f_{\mathrm{2}}(x)$ has the same form as
a Kappa-distribution which is frequently interpreted as a
consequence of acceleration mechanisms in the solar corona
(Leubner 2004).

Let us find values of the distribution parameters $\lambda$ (see
Eq.~1) and q (see Eq.~2) from the ASCA data for the HCG 62 group
and from the Beppo-SAX data for the Centaurus cluster:

\begin{enumerate}
\item The HCG 62 group. Fukazawa et al. (2001) using the observed luminosity ratio of
the non-thermal and thermal X-ray continuum components, and the
non-thermal bremsstrahlung model (Kempner \& Sarazin 2000) estimated
that the non-thermal electron population is 6\% of the thermal
electron population (the non-thermal electron energy density is 25\%
of the thermal electron energy density). We obtain the values of
$\lambda=0.28$ and $q=0.97$ by integrating over the electron spectra
$f_{\mathrm{1}}(x)$ and $f_{\mathrm{2}}(x)$ respectively.

\item The Centaurus cluster. The total luminosity of the
non-thermal component in the 20-200 keV band was calculated by
Molendi et al. (2002) from a Beppo-SAX observation. Molendi et al.
(2002) considered the bremsstrahlung model as one of the
possibilities to explain the hard X-ray excess. The luminosity in
the 2-10 keV band was calculated by Fabian et al. (2005).
From the observational data we derive that the non-thermal electron
population is 5\% of the thermal electron population (the
non-thermal electron energy density is 21\% of the thermal electron
energy density) and obtain the values of $\lambda=0.25$ and
$q=0.975$ by integrating over the electron spectra
$f_{\mathrm{1}}(x)$ and $f_{\mathrm{2}}(x)$ respectively.
\end{enumerate}
Note that supra-thermal electron populations that are 4\% and 8\%
of the thermal electrons were also proposed by Sarazin \& Kempner
(2000) for the Coma and Abell 2199 clusters.

Evidence for non-thermal X-rays in galaxy clusters is still
controversial (e.g. Rossetti \& Molendi 2004; Kitaguchi et al.
2007). The Suzaku observation of the Coma cluster does not provide
evidence for non-thermal excess in the central region of the Coma
cluster (Wik et al. 2009). An analysis of Suzaku XIS and HXD
measurements of HCG 62 resulted in an upper limit on non-thermal
emission (Tokoi et al. 2008), but at a level which does not
exclude the ASCA result.


The influence of the derived high subrelativistic electron
populations on the metal abundance estimates will be considered in
Sect.~2.2.

\subsection{The influence of high energy electron populations on metal abundance estimates.}

We now show that the effect of high energy subrelativistic
electrons on hydrogen-like and helium-like emission lines can be
significant. A generalized approach to calculate the emissivity in
hydrogen-like and helium-like spectral (iron) lines for a modified
Maxwellian electron distribution was given by Prokhorov et
al. (2009). In this section we propose to study the sum of the
H-like and He-like line volume emissivities (in units of photons
cm$^{-3}$ s$^{-1}$) instead of the line volume emissivity ratio.

The sum of the H-like and He-like line volume emissivities for a
chemical element of atomic number Z can be written as
\begin{equation}
\varepsilon_{\mathrm{Z}}=n_{\mathrm{e}} n_{\mathrm{H}}
A_{\mathrm{Z}} \times (\xi_{\mathrm{Z-2}} Q_{\mathrm{Z-2}} +
\xi_{\mathrm{Z-1}} Q_{\mathrm{Z-1}} +
\xi_{\mathrm{Z-1}}\alpha_{\mathrm{Z-2}} +
\xi_{\mathrm{Z}}\alpha_{\mathrm{Z-1}}) \label{emissivity}
\end{equation}
where $n_{\mathrm{e}}$ is the electron number density,
$n_{\mathrm{H}}$ is the H ionic number density, $A_{\mathrm{Z}}$ is
the abundance of the considered chemical element,
$\xi_{\mathrm{Z-2}}$ and $\xi_{\mathrm{Z-1}}$ are the ionic
fractions of He-like and H-like ions respectively,
$Q_{\mathrm{Z-2}}$ and $Q_{\mathrm{Z-1}}$ are the impact excitation
rate coefficients, and $\alpha_{\mathrm{Z-2}}$ and
$\alpha_{\mathrm{Z-1}}$ are the rate coefficients for the
contribution from radiative recombination to the spectral lines:
He-like triplet and H-like doublet.

Let $U$ denote the reduced expression for the sum of emissivities
$\varepsilon_{\mathrm{Z}}$ defined as:
\begin{equation}
U=\frac{\xi_{\mathrm{Z-2}} Q_{\mathrm{Z-2}} + \xi_{\mathrm{Z-1}}
Q_{\mathrm{Z-1}} + \xi_{\mathrm{Z-1}}\alpha_{\mathrm{Z-2}} +
\xi_{\mathrm{Z}}\alpha_{\mathrm{Z-1}}}{\Gamma}
\end{equation}
where $\Gamma=Z^{-4} \pi a^2_{\mathrm{0}} \sqrt{I_{\mathrm{Z}}/
m_{\mathrm{e}}}$ corresponds to the characteristic rate
coefficient value, $m_{\mathrm{e}}$ is the electron mass,
$a_{\mathrm{0}}$ is the Bohr radius, and $I_{\mathrm{Z}}$ is the
K-shell ionization potential.

For the sake of clarity, we consider in more detail the electron
distribution $f_{\mathrm{1}}(x)$ because of the distinct
non-thermal power-law component at $x\geq3$. In this case the
non-thermal electrons have energies higher than 3kT, which
corresponds to the energies $E_{\mathrm{HCG 62}}=3$ keV and
$E_{\mathrm{A3526}}=10.5$ keV for the HCG 62 group  and the
Centaurus cluster (Abell 3526). Ionic fractions are very sensitive
to the electron population with energies higher than the K-shell
potential $I_{\mathrm{Z}}$, therefore ions of argon
$I_{\mathrm{Z=18}}=4.4$ keV and iron $I_{\mathrm{Z=26}}=9.2$ keV
are promising for our analysis of electron distributions in the
HCG 62 group and in the Centaurus cluster respectively.

In the analysis of the reduced emissivity U we use the method
which was proposed by Prokhorov et al. (2009) to take into account
the influence of the high energy subrelativistic electron
population on the He-like and H-like line emissivities. All the
necessary coefficients to calculate the direct ionization cross
sections are taken from Arnaud \& Rothenflug (1985), the radiative
recombination rates are taken from Verner \& Ferland (1996), and
the dielectronic recombination rates are taken from Mazzotta et
al. (1998). Note that the fraction of Li-like ions of Ar at
temperatures kT$\geq$1 keV is less than 5\% and the fraction of
Li-like ions of Fe at temperatures kT$\geq$3.5 keV is less than
12\% (e.g. Mazzotta et al. 1998). We have included the Li-like ion
fractions in the analysis of the ionization balance.

In Fig.~1 we compare the reduced argon emissivity for the
Maxwellian electron distribution and for the modified Maxwellian
electron distribution $f_{\mathrm{1}}(x)$, with a fraction of high
energy subrelativistic electrons equal to 6\% as in the HCG 62
group.

\begin{figure}[h]
\centering
\includegraphics[angle=0, width=8cm]{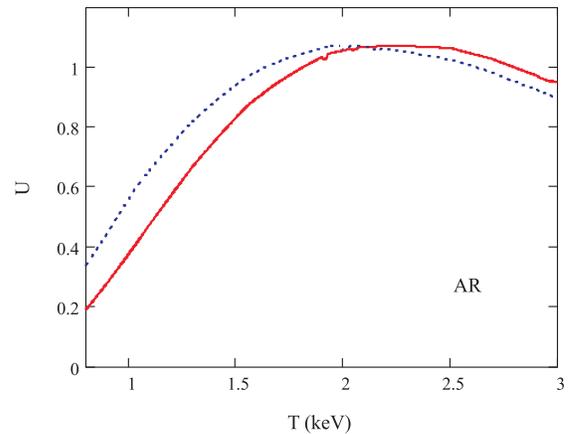}
\caption{Reduced argon emissivity $U$ for a Maxwellian electron
distribution (solid line) and for a modified Maxwellian
distribution $f_{\mathrm{1}}(x)$ (dashed line) for the HCG 62
group.}
\end{figure}

For the HCG 62 group (kT=1 keV) the reduced emissivity $U$ for a
modified Maxwellian distribution $f_{\mathrm{1}}(x)$ increases by
$\approx49\%$ with respect to the case of a Maxwellian
distribution. Such an increase of the reduced emissivity U
corresponds to a decrease of the argon abundance $A_{Z=18}$ for a
constant value of $\varepsilon_{\mathrm{Z=18}}$ (see
Eq.~\ref{emissivity}). For a modified Maxwellian distribution
$f_{\mathrm{1}}(x)$ the argon abundance decreases by $\approx33\%$
with respect to the case of a Maxwellian distribution. The
decrement of the argon abundance for a modified Maxwellian
distribution $f_{\mathrm{2}}(x)$ is $27\%$.

Abundances of Mg, Si, S and Fe in the HCG 62 group were calculated
from Suzaku data by Tokoi et al. (2008), but the expected Ar
abundance is 4.5 times smaller than the S abundance (Anders \&
Grevesse 1989) and it is more difficult to detect Ar lines. The
predicted Ar abundance decrease is a tool to test the
bremsstrahlung interpretation of the hard X-ray tail in HCG 62.

The Centaurus cluster is another interesting object to study. In
Fig.~2 we compare the reduced iron emissivity for a Maxwellian
electron distribution and for a modified Maxwellian electron
distribution $f_{\mathrm{1}}(x)$ with a fraction of high energy
subrelativistic electrons equal to 5\%, as in the Centaurus
cluster.

\begin{figure}[t]
\centering
\includegraphics[angle=0, width=8cm]{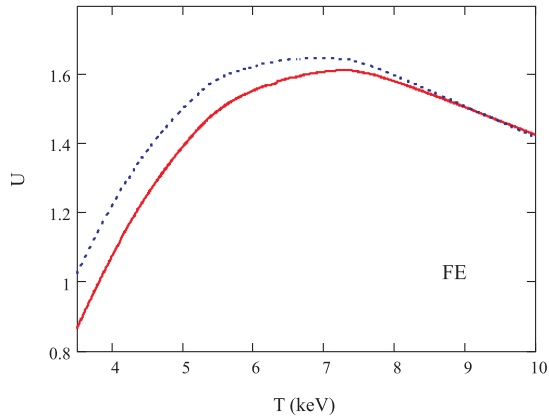}
\caption{Reduced iron emissivity $U$ for a Maxwellian electron
distribution (solid line) and for a modified Maxwellian
distribution $f_{\mathrm{1}}(x)$ (dashed line) for the Centaurus
cluster.}
\end{figure}

We found that the iron abundance for the modified Maxwellian
distributions $f_{\mathrm{1}}(x)$ and $f_{\mathrm{2}}(x)$
decreases by $\approx15\%$ and $\approx 13\%$ respectively with
respect to the case of a Maxwellian distribution.

We also calculated changes in the abundance estimates for the
chemical elements closest in atomic numbers to Ar and Fe and found
that: 1) for HCG 62, the Si abundance increases by 3\%, and the S
abundance decreases by 14\%;
2) for the Centaurus cluster, the Ar abundance increases by 2\%,
and the Ca abundance increases by 0.5\%.

High energy subrelativistic electrons lead to a higher apparent
temperature. If the gas temperature is smaller than the
temperature at which the reduced emissivity U of the chemical
element has a maximum value, then the abundance estimate decreases
because of an increase of the reduced emissivity with the
temperature at these temperatures. However, if the gas
temperature is higher than the temperature at which the reduced
emissivity U of the chemical element has its maximum value, then the
abundance estimate increases.

We now demonstrate how the argon and iron abundances inferred from
X-ray observations yield important constraints on the fraction of
high energy electrons. For this purpose, synthetic clusters with
temperatures of 1 and 3.5 keV (as in the HCG 62 group and in the
Centaurus cluster) and an electron distribution function
$f_{\mathrm{1}}(x)$ are considered. The dependences of both argon
and iron abundance ratios for a modified Maxwellian distribution
and for a Maxwellian distribution on the fraction of high energy
subrelativistic electrons are shown in Fig.~3.

\begin{figure}[ht]
\centering
\includegraphics[angle=0, width=8cm]{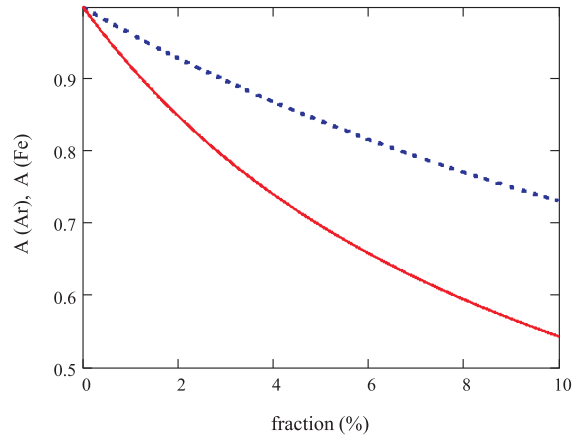}
\caption{The solid (dashed) line shows dependence of the ratio of
the argon (iron) abundances for a modified Maxwellian distribution
and for a Maxwellian distribution on the fraction of high energy
electrons.}
\end{figure}

We conclude that high energy electron populations can affect
derived metal abundances for the HCG 62 group and the Centaurus
cluster.

\section{Discussion}

It can be supposed that, in addition to the
bremsstrahlung-emitting thermal ICM and synchrotron-emitting,
relativistic, non-thermal electrons, a high energy subrelativistic
population of electrons exists which emits the hard X-ray excess
as bremsstrahlung.


There are three possible origins for high energy subrelativistic
populations: non-thermal (Sarazin \& Kempner 2000), quasi-thermal
(Blasi 2000; Dogiel 2000; Liang et al. 2002; Dogiel et al. 2007;
Wolfe \& Melia 2008), and thermal in the framework of the
non-extensive thermo-statistics (Hansen 2005). Since line
emissivities depend on the fraction of high energy electrons (e.g.
Prokhorov et al. 2009) and do not depend on the origin of these
electrons, we can use the electron distribution $f_{\mathrm{1}}(x)$
to calculate line emissivities in the cases of the non-thermal and
quasi-thermal electron origins.

Petrosian (2001) estimated the yield in non-thermal bremsstrahlung
photons $Y\sim (dE/dt)_{\mathrm{br}}/(dE/dt)_{\mathrm{c}}\sim
10^{-5}$. Here $(dE/dt)_{\mathrm{br}}/(dE/dt)_{\mathrm{c}}$ is the
ratio of bremsstrahlung to Coulomb losses of non-thermal
electrons. Then for a hard X-ray flux $F_{\mathrm{x}}\sim 10^{43}$
erg/s a large amount of energy of the non-thermal electrons
$F_{\mathrm{e}}\sim F_{\mathrm{x}}/Y\sim 10^{48}$ erg/s is
transmitted to the background plasma. As a result the ICM should
be heated above its observable temperature in less than 10 Myr
(Petrosian 2001, Wolfe \& Melia 2006).

It was, however, shown by Liang et al. (2002) and Dogiel et al.
(2007) that a quasi-thermal electron population might overcome
this difficulty via a higher radiative efficiency (and therefore a
longer overheating time, but see Petrosian \& East 2008). The
energy supply necessary to produce the observed hard X-ray flux by
quasi-thermal electrons is at least one or two orders of magnitude
smaller (Dogiel et al. 2007) than derived form the assumption of
non-thermal origin of emitting electrons. Wolfe \& Melia (2008)
have also considered a quasi-thermal electron distribution to fit
hard X-ray emission, but rather than requiring a second-order
Fermi acceleration to produce the quasi-thermal electrons, they
assumed quasi-thermal electrons are produced via collisions with
non-thermal protons.

\section{Conclusions}

We have shown in this paper that the metal abundance estimates
depend on the presence of high energy subrelativistic electrons
proposed to account for measurements of hard X-ray excess emission
from galaxy groups and clusters. Due to the impact of these
energetic electron populations, the Ar abundance estimate in the
HCG 62 group and the Fe abundance estimate in the Centaurus
cluster significantly decrease by $\approx30$ and $\approx15\%$,
respectively.

These decreases in the Ar and Fe abundance estimates are determined
by the high energy subrelativistic electron fractions (6\% for the
HCG 62 group and 5\% for the Centaurus cluster) which are comparable
to the thermal electron fractions with energies higher than the
K-shell ionization potentials of Ar and Fe, respectively.

The influence of the high energy subrelativistic electron
populations on the abundance estimates is measurable with current
instruments. Therefore this probe is more efficient for detecting
the high energy subrelativistic electron populations than that
based on the Doppler broadening of the spectral lines proposed by
Hansen (2005), which requires very high energy resolution, or than
that based on the flux ratio of the emission lines (Prokhorov et
al. 2009), which requires higher sensitivity instruments.

Other possibilities to produce the change in the metal abundance
estimates in galaxy clusters are the effect of resonant scattering
(Gilfanov et al. 1987) and the presence of multiphase hot gas -
two temperature model (Buote \& Fabian 1998, Buote 2000).

The effect of resonant scattering causes the decrement of the
FeXXV line at 6.7 keV, and, therefore, the decrement of the flux
ratio of the iron lines FeXXV/FeXXVI. A decrement of the Fe
abundance is then produced, as in the case of a high energy
subrelativistic electron population. To separate the effects of
resonant scattering and the high energy subrelativistic population
influence, the SZ effect from a high energy subrelativistic
population can be analyzed. Following the method of Colafrancesco
et al. (2009), we calculated the value of the slope of the SZ
effect in the Centaurus cluster. We obtained the value of the
slope $S\approx0.033$ for both electron distributions
$f_{\mathrm{1}}(x)$ and $f_{\mathrm{2}}(x)$ and the value of the
slope $S\approx0.028$ for a Maxwellian spectrum. Since the slope
is equal to $S\approx4.25\times kT/(m_{\mathrm{e}} c^2)$ for a
Maxwellian electron spectrum without a high energy subrelativistic
electron population (Colafrancesco et al. 2009), the value of the
slope of $S=0.033$ corresponds to that at an effective temperature
$kT=4.5$ keV, which is higher than the temperature $kT=3.5$ keV
observed with Beppo-SAX.

Buote (2000) analyzed the ASCA data for the HCG 62 group, the same
data as analyzed by Nakazawa et al. (2007), and fitted the spectrum
in the frame of a two-temperature model with temperatures
$kT_{\mathrm{1}}=0.7$ keV and $kT_{\mathrm{2}}=1.4$ keV. Nakazawa
et al. (2007) showed that the spectrum of the HCG 62 group is well
reproduced by the two-temperature model of Buote (2000) in the
energy range below $\sim 4$ keV, but a fit of the full spectrum
requires a thermal component with an unrealistically high
temperature of $\sim17.5$ keV.

We have shown that high energy electron populations can affect
derived metal abundances in galaxy groups and clusters and in the
solar corona. Therefore, metal abundances are a promising tool for
an analysis of the high energy subrelativistic electron component.

\begin{acknowledgement}
I am grateful to Florence Durret, Joseph Silk, William Forman,
Vladimir Dogiel and Eugene Churazov for valuable suggestions and
discussions and thank the referee for very useful comments.
\end{acknowledgement}

\begin{appendix}

\section{Oxygen abundance drop in the solar corona}

Often the particle distribution functions in space plasmas, e.g.
solar corona plasma, are observed to be quasi-Maxwellian at the
mean thermal energies, while they have non-Maxwellian
supra-thermal tails at higher energies (e.g. Porquet et al. 2001).
A Kappa-distribution is very convenient to model these particle
distribution functions, since it fits both the thermal and
supra-thermal parts of the observed energy spectra (for a review,
see Leubner 2004).

The ionization and excitation rates for elements C, O, Fe for a
Kappa-distribution of electrons in the solar corona were studied
by Owocki \& Scudder (1983) and Dzifcakova \& Kulinova (2001,
2003).

We now demonstrate how the oxygen abundance inferred from X-ray
observations yields important constraints on the fraction of high
energy electron populations. For this purpose, a plasma with
temperature $kT = 0.1$ keV (as in the solar corona) and an
electron distribution function $f_{\mathrm{2}}(x)$ are considered.
The dependence of the ratio $A(O)$ of the oxygen abundances for a
modified Maxwellian distribution and for a Maxwellian distribution
on the fraction of high energy subrelativistic electrons is shown
in Fig.~A.1. Due to the impact of high energy subrelativistic
electrons, the oxygen abundance estimate in the solar corona
significantly decreases. Even if the fraction of high energy
electrons is only 1\% the drop of the oxygen abundance is about
30\%.

\begin{figure}[ht]
\centering
\includegraphics[angle=0, width=8cm]{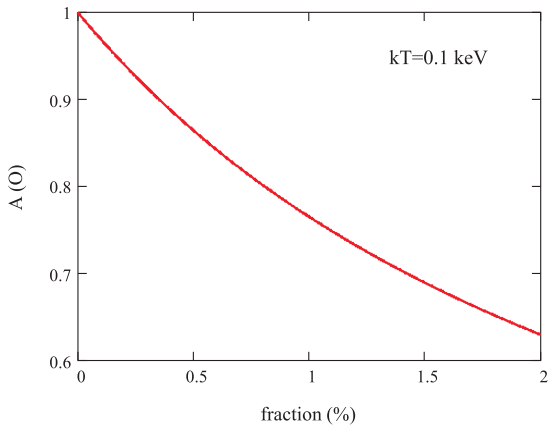}
\caption{Dependence of the ratio of the oxygen abundances for a
modified Maxwellian distribution and for a Maxwellian distribution
on the fraction of high energy electrons.}
\end{figure}
\end{appendix}


\begin{thebibliography}{99}
\bibitem{Anders 1989}
Anders, E., Grevesse, N. 1989, Geochim. Cosmochim. Acta, 53, 197
\bibitem{Arnaud 1985}
Arnaud, M., Rothenflug, R. 1985, A\&ASS, 60, 425
\bibitem{Blasi 2000a}
Blasi, P. 2000, ApJL, 532, L9
\bibitem{Blasi 2000b}
Blasi, P., Olinto, A. V., Stebbins, A. 2000, ApJ, 535, L71
\bibitem{Brunetti 2003}
Brunetti, G. 2003, in Matter and Energy in Clusters of Galaxies,
ASP Conference Proceedings, 301, 349
\bibitem{Boute 1998}
Buote, D. A., Fabian, A. C. 1998, MNRAS, 296, 977
\bibitem{Buote 2000}
Buote, D. A. 2000, MNRAS, 311, 176
\bibitem{Clarke 2001}
Clarke, T. E., Kronberg, P. P., B\"{o}hringer, H. 2001, ApJ, 547,
L111
\bibitem{Colafrancesco 2009}
Colafrancesco, S., Prokhorov, D. A., Dogiel, V. A. 2009, A\&A, 494,
1
\bibitem{Dogiel 2000}
Dogiel, V. A. 2000, A\&A, 357, 66
\bibitem{Dogiel 2007}
Dogiel, V. A., Colafrancesco, S., Ko, C. M. et al. 2007, 461, 433
\bibitem{Dzifcakova 2001}
Dzifcakova, E., Kulinova, A. 2001, Solar Physics, 203, 53
\bibitem{Dzifcakova 2003}
Dzifcakova, E., Kulinova, A. 2003, Solar Physics, 218, 41
\bibitem{Fabian 2005}
Fabian, A. C., Sanders, J. S., Taylor, G. B.,  Allen, S. W. 2005,
MNRAS, 360, 20
\bibitem{Ferrari 2008}
Ferrari, C., Govoni, F., Schindler, S. et al. 2008, SSRv, 134, 93
\bibitem{Fukazawa 2001}
Fukazawa, Y., Nakazawa, K., Isobe, N. et al. 2001, ApJ, 546, L87
\bibitem{Fusco-Femiano 1999}
Fusco-Femiano, R., Dal Fiume, D., Feretti, L. et al. 1999, ApJ, 513,
L21
\bibitem{Fusco-Femiano 2004}
Fusco-Femiano, R., Orlandini, M., Brunetti, G. et al. 2004, ApJ,
602, L73
\bibitem{Gilfanov 1987}
Gilfanov, M. R., Sunyaev, R. A., Churazov, E. M. 1987, Sov. Astron.
Lett., 13, 233
\bibitem{Hansen 2005}
Hansen, S. 2005, New Astronomy, 10, 371
\bibitem{Kaastra 1999}
Kaastra, J. S., Lieu, R., Mittaz, J. P. D. et al. 1999, ApJ, 519,
L119
\bibitem{Kaastra 2009}
Kaastra, J. S., Bykov, A. M., Werner, N. 2009, A\&A, in press,
astro-ph/0905.4802
\bibitem{Kempner 2000}
Kempner, J. C., Sarazin, C. L. 2000, ApJ, 530, 282
\bibitem{Kitaguchi 2007}
Kitaguchi, T., Makishima, K., Nakazawa, K. et al. 2007, in Proc. of
`The Extreme Universe in the Suzaku era'
\bibitem{Leubner 2004}
Leubner, M. P. 2004, ApJ, 604,469
\bibitem{Liang 2002}
Liang, H., Dogiel, V., Birkinshaw, M. 2002, MNRAS, 337, 567
\bibitem{Mazzotta 1998}
Mazzotta, P., Mazzitelli, G., Colafrancesco, S.et  al. 1998,
A\&AS, 133, 403
\bibitem{Mewe 1981}
Mewe, R., Gronenschild, E. H. B. M. 1981, ApSS, 45, 11
\bibitem{Molendi 2002}
Molendi, S., De Grandi, S., Guainazzi 2002, A\&A, 392, 13
\bibitem{Nakazawa 2007}
Nakazawa, K., Makishima, K., Fukazawa, Y. 2007, PASJ, 59, 167
\bibitem{Owocki 1983}
Owocki, S. P., Scudder, J. D. 1983, ApJ, 270, 758
\bibitem{Petrosian 2001}
Petrosian, V. 2001, ApJ, 557, 560
\bibitem{Petrosian 2008a}
Petrosian, V., Bykov, A. M., Rephaeli, Y. 2008, SSRv, 134, 191
\bibitem{Petrosian 2008b}
Petrosian, V., East, W. 2008, ApJ, 682, 175
\bibitem{Porquet 2001}
Porquet, D., Arnaud, M., Decourchelle, A. 2001, A\&A, 373, 1110
\bibitem{Prokhorov 2009}
Prokhorov, D. A., Durret, F., Dogiel, V. A., Colafrancesco, S. 2009,
A\&A, 496, 25
\bibitem{Rephaeli 2008}
Rephaeli, Y., Nevalainen, J., Ohashi, T., Bykov, A. M. 2008, SSRv,
134, 71
\bibitem{Rossetti 2004}
Rossetti, S., Molendi, S. 2004, A\&A, 414, 41
\bibitem{Sarazin 1986}
Sarazin, C. L. 1986, Rev. Mod. Phys., 58, 1
\bibitem{Sarazin 1998}
Sarazin, C. L., Lieu, R. 1998, ApJ, 494, L177
\bibitem{Sarazin 2000}
Sarazin, C. L., Kempner, J. C. 2000, ApJ, 533, 73
\bibitem{Shimon 2002}
Shimon, M., Rephaeli, Y. 2002, ApJ, 575, 12
\bibitem{Silva 1998}
Silva, R., Plastino, A. R., Lima, J. A. S. 1998, Physics Letters
A, 249, 401
\bibitem{Tokoi 2008}
Tokoi, K., Sato, K., Ishisaki, Y. et al. 2008, PASJ, 60, 317
\bibitem{Tsallis 1988}
Tsallis, C. 1988, J. Stat. Phys., 52, 479
\bibitem{Tsallis 1999}
Tsallis, C. 1999, BrJPh, 29, 1
\bibitem{Verner 1996}
Verner, D. A., Ferland, G. J. 1996, ApJS, 103, 467
\bibitem{Werner 2008}
Werner, N., Durret, F., Ohashi, T. et al. 2008, Space Sci. Rev.,
134, 337
\bibitem{Wik 2009}
Wik, D. R., Sarazin, C. L., Finoguenov, A. et al. 2009, ApJ, 696,
1700
\bibitem{Wolfe 2006}
Wolfe, B., Melia, F. 2006, ApJ, 638, 125
\bibitem{Wolfe 2008}
Wolfe, B., Melia, F. 2008, ApJ, 675, 156
\bibitem{Zeldovich 1969}
Zel'dovich, Ya. B., Sunyaev, R. A. 1969, Ap\&SS, 4, 285
\end{thebibliography}
\end{document}